\documentclass[useAMS,usenatbib,letterpaper]{mn2e}

\usepackage{graphicx}

\title{Blasts and shocks in the disc of NGC 4258}
\author[]{J. Jim\'enez-Vicente$^{1,5}$,  E. Mediavilla$^{2,4}$, A. Castillo-Morales
$^{3}$, and E. Battaner$^{1,5}$\\
$^{1}$Dpto de F\'{\i}sica Te\'orica y del Cosmos. Universidad de Granada, Spain.\\
$^{2}$Instituto de Astrof\'{\i}sica de Canarias. Tenerife. Spain.\\
$^{3}$Dpto de Astrof\'{\i}sica y C.C. de la Atm\'osfera. Univ. Complutense de Madrid, Spain.\\
$^{4}$Dpto. de Astrof\'{\i}sica. Universidad de La Laguna, Spain\\
$^{5}$Instituto ``Carlos I'' de F\'{\i}sica Te\'orica y Computacional. Universidad de Granada, Spain}
\begin{document}


\maketitle
\label{firstpage}
\begin{abstract}

We present integral field spectroscopic observations of the central 
region of the active galaxy NGC 4258
obtained with the fibre IFU system INTEGRAL. 
We have been able to detect cold neutral gas by means of the interstellar NaD
doublet absorption and to trace its distribution and kinematics with respect 
to the underlying disc. 
The neutral gas is blue-shifted with projected velocities in the 120--370 
km/s range.
We have also detected peculiar kinematics in part of the ionized gas
in this region by means of a careful kinematic decomposition. 
The bipolar spatial distribution of the
broader component is roughly coincident with the morphology of the
X-ray diffuse emission. The kinematics of this gas can be explained in terms of
expansion at very high (projected) velocities of up to 300 km/s. 
The observations also reveal
the existence of a strip of neutral gas, parallel to
the major kinematic axis, that is nearly
coincident with a region of very high [SII]/H$\alpha$ ratio 
tracing the shocked gas. 
Our observations are consistent with the jet model presented by \cite{wilsonetal01} 
in which a cocoon originating from the nuclear jet is shocking the
gas in the galaxy disc. Alternatively, our observations are also consistent
 with the bipolar hypershell model of \cite{Sofue80} and \cite{SofueandVogler01}. On balance, 
we prefer the latter model as the most likely explanation for the puzzling features of this peculiar object.


\end{abstract}
\section{Introduction}
\label{intro}

The nearby spiral galaxy NGC 4258 (M106) is a particularly interesting galaxy. 
It presents two peculiarities 
that have been the cause of many studies of this object: 
the confirmation of the existence at its centre of a massive black hole, 
detected in 1995 through
measurements of high speed motions in the water masers (\cite{miyoshi95}), 
and the existence of the mysterious {\em anomalous arms} discovered in the optical by \cite{courtes61} 
and in radio by \cite{vdkruit72}. As we shall see below, it is indeed not 
unlikely that both phenomena are closely related.

These {\em anomalous arms} are two large arm-like gaseous structures of
uncertain origin that extend through most of the visible disc of the galaxy.
There have been several possible explanations for the origin of these arms, 
most of which can be traced back to three distinct possibilities:
(a) expelled gas from the nucleus in the plane, either in a ballistic 
(\cite{vdkruit72}, \cite{valbada78}, \cite{valbadavhulst82}) or a jet model 
(\cite{fordetal86}, \cite{ceciletal92}), (b) a bar-induced shock front (\cite{coxdownes96}), 
and (c) gas expelled out of the plane either in a jet, hypershell, 
etc. (\cite{Sofue80}, \cite{Sanders82}, \cite{SofueandVogler01}). 
There are indeed many observational features that have to be explained 
by any suitable model (see for example the excellent works by 
\cite{ceciletal00} and \cite{wilsonetal01} including a plethora of optical, radio and X-ray data).

The morphology of the  anomalous arms can
be greatly influenced by projection effects. Although it is pretty well established 
that the inner parts of these arms are in the disc plane, it is not known for certain 
whether the arms are three dimensional structures seen in projection (and enhanced 
by a limb-brightening effect), or if they are indeed mostly contained in a 
plane (and if this is the case, the orientation of this plane). On the other hand, 
the highly distorted kinematics that appear in the innermost region of the 
galaxy also imposes restrictions on the possible scenarios.

The galaxy NGC 4258 has an inclination of about 60--64$^\circ$, and a
 position angle of 146--160$^\circ$ (cf. \cite{jarrettetal03}, \cite{ceciletal92}).
 The strong absorption below the major axis shows clearly that the region located 
SW of the major axis is the near side of the galaxy, the one on the NE being the far side.

The most relevant observational facts are: a) the anomalous arms are made up
of hot gas that has been shock excited; b) the radio emission of the innermost region 
shows a clear jet (nearly in the N--S direction), perpendicular to the maser 
disc around the black hole (The maser disc is actually tipped down some 
8$^\circ$ from edge on [see \cite{herrnsteinetal97}], meaning that the northern 
jet is pointing toward us while the southern one is pointing away from us);
c) the NW (SE) arm lies behind (in front of) the 
galaxy disc; and d) the inner section of the anomalous arms is roughly straight
and it is roughly aligned with the major axis of the galaxy.

From the several scenarios that have been proposed to explain the observational facts, the 
model that seems to be favoured by the latest observations in X-rays
(\cite{ylwr07}) is the one proposed by {\cite{wilsonetal01}) in which a
jet moving away from the disc induces strong shocks in the disc plane
giving rise to the anomalous arms.

We have made new observations of the inner region of NGC 4258 with an 
integral field spectrograph. These new observations will provide new kinematic 
and morphological information that will permit us to put the available models to the test.

\section{Observations, and data reduction and analysis}
\label{obsandred}

The data analysed in this paper were obtained on 2002 March 16 at the 
Observatorio del Roque de los
Muchachos on the island of La Palma with the INTEGRAL fibre system 
~\citep{arribas98} in combination with the WYFFOS fibre 
spectrograph ~\citep{bingham94} 
at the William Herschel Telescope. Weather conditions during observations 
were fairly good, 
with an average
seeing of 1.3 $\arcsec$. The data were obtained with 
INTEGRAL standard bundles SB3 and SB2, and
the WYFFOS spectrograph was equipped with a 1200 groove $\mathrm{ mm^{-1}}$ 
grating 
centred on 6247 $\mathrm\AA$. With these settings, the field of view is 
$33.6\arcsec \times 29.4\arcsec$ and  $16.0\arcsec \times 12.3 \arcsec $ 
respectively for the SB3 and SB2 bundles. 
The spectral resolution is 4.8 $\mathrm\AA$ (R $\approx$ 1300) for the 
SB3 bundle 
and 2.8 $\mathrm\AA$ (R $\approx$ 2200) for the SB2 bundle. 
The observed spectral range (5600--6850 $\mathrm\AA$) contains a few strong interstellar 
emission lines ($\mathrm H\alpha$, [NII]$\lambda\lambda6548, 6584$, 
[SII]$\lambda\lambda6716, 6731$), as well as several stellar absorption lines (NaD doublet
and weaker, mostly blended FeI and CaI lines). 
Three exposures of 1200 s were taken for each fibre bundle.
Standard data reduction for 2D spectroscopic data (bias
correction, flat-fielding, extraction of spectra, wavelength calibration, sky
subtraction, etc.) was performed (e.g.\ \cite{Arribasetal91}, \cite{Mediavillaetal92}).
The final wavelength calibration
has uncertainties of 0.15 $\mathrm{\AA}$ or 7 km/s at 6500 $\mathrm{\AA}$.

In order to separate the stellar and interstellar contribution to the spectra, 
each spectrum is fitted using a synthetic stellar population as a template.
The regions in the spectrum which are obviously contaminated by interstellar
emission and/or absorption (namely $\mathrm H\alpha$, 
[NII]$\lambda\lambda6548, 6584$, [SII]$\lambda\lambda6716, 6731$ and NaD) 
are not included in the fit.
After some testing, we have finally used a population of solar metallicity with 
an age of 1 Gyr from the synthetic population library by~\citet{delgado05}, 
which covers the observed wavelength
range at a very high spectral resolution (0.3 \AA). 
Using populations with a slightly different age
and/or metallicity do not seriously affect the results. We would like to 
point out that we do not intend a thorough analysis of the stellar populations
in this region, but just to be able to subtract the stellar contribution from
the observed spectra. As a by-product we obtain a stellar velocity, velocity
dispersion and
a global strength for the stellar lines at each fibre position. 
More details on this procedure and the general reduction
process can be found in ~\citep{Castilloetal07} and \cite{Jimenezetal07}.




Maps for several spectral features are built from the extracted spectra 
by mapping the extracted 
values at the fibre positions to a regular grid. 
The final images have 33 $\times$ 29 pixels (with a size of 0.95 arcsec/pixel) 
and 45 $\times$ 34 pixels (with a size of 0.35 arcsec/pixel) for the SB3 and SB2 bundles respectively. 
The resulting
FOV are theferore of $31.35''\times 27.55''$ for the SB3 and $15.75''\times 11.9''$ for
the SB2.
Finally, astrometry was calculated for the maps by using the position
of the nucleus and bright HII regions in the FOV.

\section{Morphology and kinematics}
\label{morphokin}

The optical images of this galaxy show very clearly that heavy 
absorption is taking place in the (near side) SW part of the galaxy (see for 
example fig. 6 in \cite{ceciletal00}).
It is very important to bear this fact in mind when interpreting the 
observations presented in this paper, particularly for the different 
effect it will have on
the emission and absorption lines.

\begin{figure}
   \centering
   \includegraphics[angle=0,width=8.5cm, clip=true]{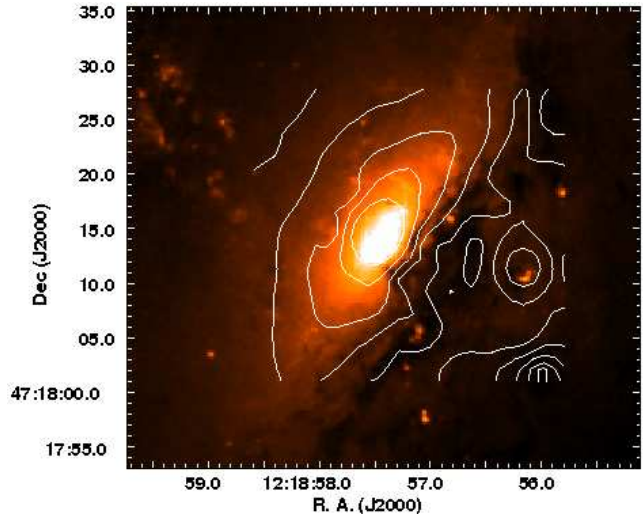}
   \caption{Image of the inner region obtained with the F658N filter 
with the ACS on the HST. Overplotted contours show our observations with 
the same (synthetic) filter.}\label{ha_and_hst}
\end{figure}

\begin{figure}
   \centering
   \includegraphics[angle=0,width=8.5cm, clip=true]{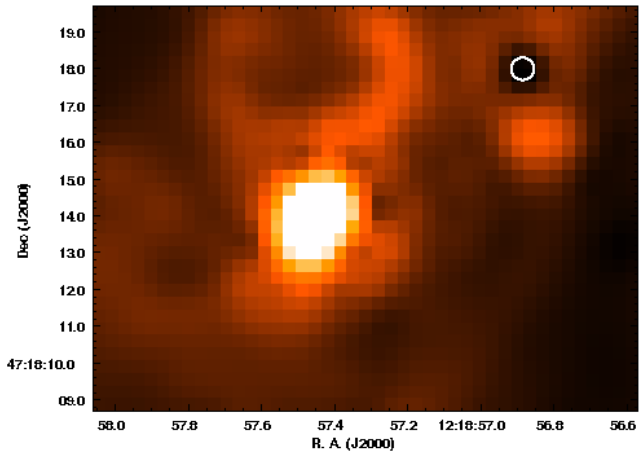}
   \caption{Continuum-free $\mathrm H\alpha$ image of the innermost region 
of NGC 4258 as observed with the SB2 fibre bundle of INTEGRAL. The circle
 on the upper right marks the defective fibre \#167.}\label{ha_sb2}
\end{figure}

The observed spectra show a few interesting emission lines from 
the ISM, namely $\mathrm H\alpha$, [NII]$\lambda\lambda6548, 6584$, 
[SII]$\lambda\lambda6716, 6731$ and [OI]$\lambda\lambda6300, 6363$.

The distribution of $\mathrm H\alpha$ intensity from the SB3 and SB2 bundles 
matches exceptionally well the features seen in the HST image taken by the ACS 
with the narrow F658N filter (see Fig. \ref{ha_and_hst}). In order to compare 
both, we have generated a synthetic F658N image from our data by integrating 
the spectra with the wavelengths weighted with the transmission curve of that 
filter. The image and contours show the above-mentioned strong absorption to the SW of the nucleus.
In the (continuum-free) 
$\mathrm H\alpha$ image from the SB2 (with higher spatial resolution), 
the northern {\em nuclear loop} first detected by \cite{ceciletal00} 
stands out even 
more clearly than in images from the HST (see Fig. \ref{ha_sb2}). 
 In our continuum-free $\mathrm H\alpha$ image, the contrast between
the E and W sides of the loop seems weaker than in \cite{ceciletal00}.}

There is even a hint of a southern counterpart of this loop
in the asymmetric morphology below the nucleus, 
although severe obscuration by dust in this region makes it impossible to 
confirm.
The nature of the loop(s) remains unclear. A possible origin is that they 
are gas bubbles originating from the interaction of the jet with the dense ISM, as 
proposed by \cite{ceciletal92}. The velocity structure of the ionized gas, 
showing an asymmetry roughly along the N--S line, supports a connection with 
the jet. On the other hand, the line ratios at those locations are not indicative 
of shocks. Thus, the possibility of the loop(s) being the edges of an ionization 
cone, as proposed in \cite{ceciletal00} is also very attractive.

\begin{figure}
   \centering
   \includegraphics[angle=0,width=8.5cm, clip=true]{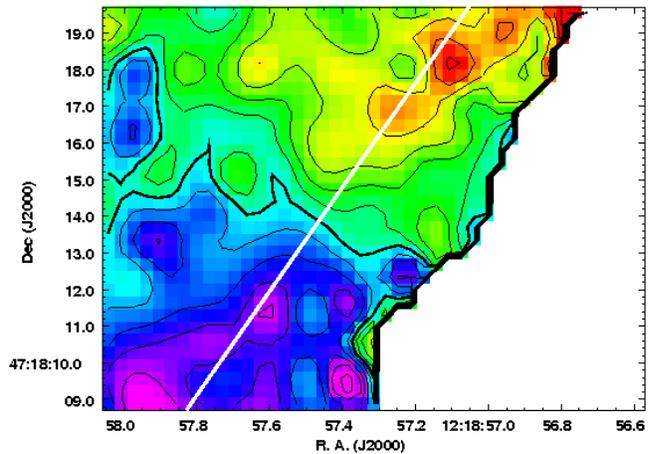}
   \caption{Star velocity field. The region in the SW is masked due to low S/N.
Major axis is indicated. Contour levels are marked every 20 km/s. 
Systemic velocity of 450 km/s is marked as a thicker contour. Blue is
 approaching and red is receding. }\label{velstars}
\end{figure}

The velocity field
in Fig. \ref{velstars}
shows the rotation pattern of the stars. A big area in the SW part of the
FOV has been masked because the low S/N in the spectra induced by the high 
extinction in this region does not allow us to 
calculate a reliable velocity. 

Even for the SB3 bundle, the FOV is 
quite small, and the velocity field of the ionized gas is too 
distorted to calculate a rotation model for the disc. We will therefore
use the velocity field for the stars to subtract the rotating pattern of
the galaxy from the gas components in order to generate residual velocity
fields. Obviously, there will be differences between the rotating pattern of 
the stars and the gas (namely the asymmetric drift). 
We do expect these differences to be small enough for our purposes and,
in any case, we will be very cautious when interpreting the residual velocity fields.

\subsection{Neutral gas distribution and kinematics}

\begin{figure}
   \centering
   \includegraphics[angle=0,width=8.5cm, clip=true]{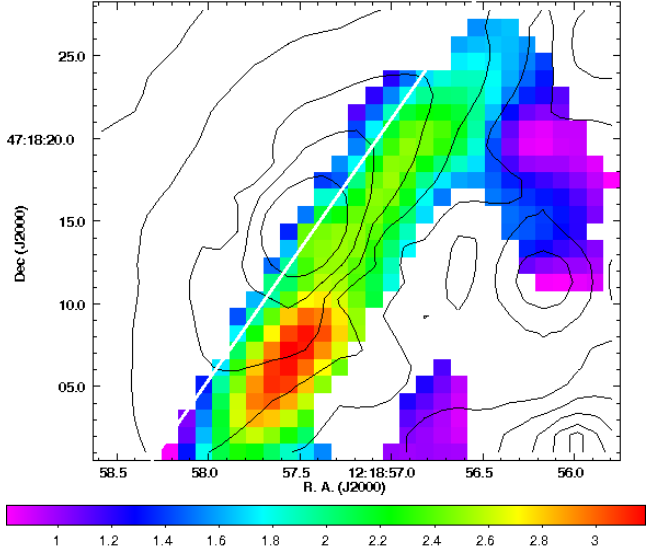}
   \caption{Map of the equivalent width of the interstellar NaD.
 Regions with an EW lower than 0.8 Angstrom have been masked out. 
Brightest spot has an EW of 3.2. The major axis is indicated. The contours 
show synthetic image with F658N filter. }\label{ewnad}
\end{figure}

\begin{figure}
   \centering
   \includegraphics[angle=0,width=8.5cm, clip=true]{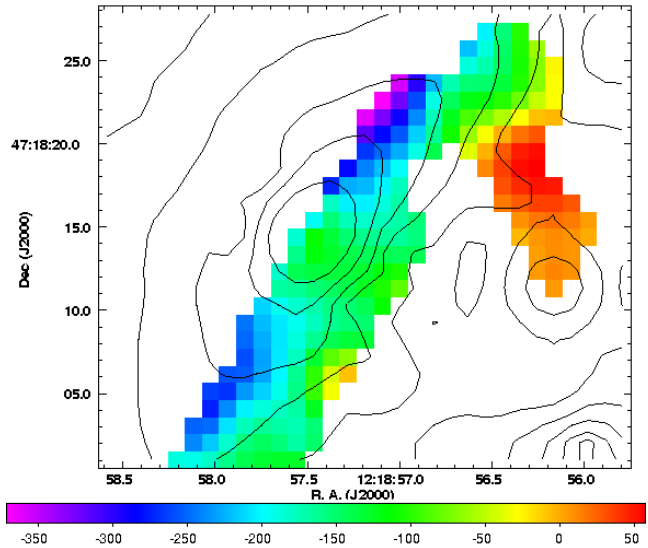}
   \caption{Map of the residual velocity field of the interstellar neutral gas.
Velocity values are shown in km/s in the colour bar at the bottom. 
Contours show synthetic image with the  F658 filter.}\label{velnamstars}
\end{figure}

In this section we address the distribution and kinematics of the interstellar neutral gas.
In Fig. \ref{ewnad} we present the distribution of the neutral gas detected by 
means of the equivalent width of the NaD doublet. In Fig. \ref{velnamstars} we 
show the residual velocities obtained by subtracting the stellar velocity map 
from the NaD velocity map. In the map of the equivalent width we can see a strip 
parallel to the major geometrical axis of the disc starting at the nucleus and
 extending some 15 arcsec to each side of the nucleus. This strip nicely traces 
the neutral gas of the near side of the galaxy that is absorbing the light coming
 from the stellar disc. Of course, this is not all the neutral gas, but only the one 
that can be detected in absorption (we cannot see the neutral gas on the far side of 
the galaxy, nor can we see the neutral gas in the more extinguished region on the 
nearest side for the same reason). Even with this obvious limitation, 
the detected distribution of cold neutral gas roughly resembles the distribution of
 molecular gas as seen in CO observations (cf.\ \cite{Sawadasatohetal07}). 

Figure \ref{velnamstars} shows the velocity of the neutral gas with respect to the
 underlying disc obtained by subtracting the stellar velocity map from the velocity 
map of the neutral gas. Again, the morphology of the map resembles the residual maps
 obtained from CO observations by \cite{Sawadasatohetal07}, but in our case the residual
 velocities are much higher.  The detected neutral gas has huge residual velocities
 in the $-120$ to $-370$ km/s range with respect to the stellar disc. Although
 \cite{Sawadasatohetal07} have proposed a warped disc with expansion to explain
 their data, we favour an alternative scenario to explain our detected 
morphology and kinematics of the neutral gas. In our opinion our data 
seem to indicate that the detected neutral gas (which is on the near side of the 
galaxy) it is most probably moving radially outwards from the galaxy centre. We
 will elaborate a bit more on this scenario later on.

\subsection{Kinematic anomalies and physical conditions of the ionized gas}

\begin{figure}
   \centering
   \includegraphics[angle=0,width=8.5cm, clip=true]{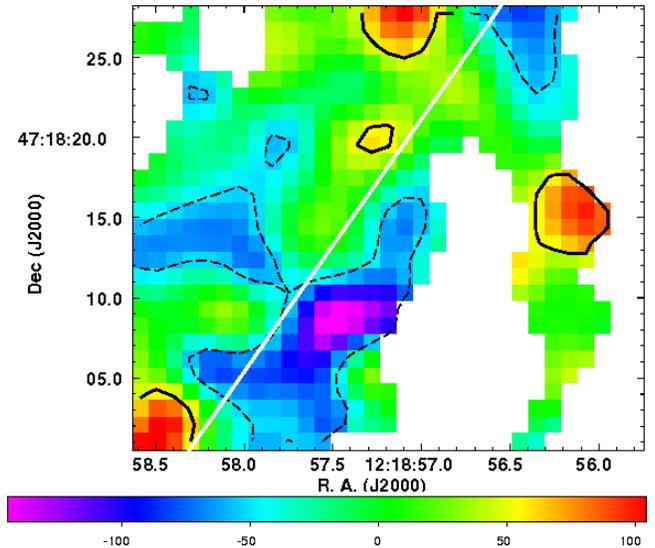}
   \caption{Map of the residual velocity field of the ionized gas.
Velocity values are shown in km/s in the colour bar at the bottom. 
The thick continuous and dashed contours show regions with residuals 
higher than 50 km/s and lower than $-50$ km/s respectively. The major axis
 is indicated by the white line.}\label{velhamstars}
\end{figure}

Comparing the velocity maps of the ionized gas (see 
Fig. \ref{velhamstars}) with the stellar velocity field, we 
can see two conspicuous protrusions: one with receding velocities 
on the approaching side, and vice versa, roughly aligned at 
PA 145$^\circ$. This strong disturbance of the velocity field is 
caused by a kinematically distinct system of ionized gas that deviates 
from simple galactic rotation.  We found direct evidence of this 
ionized gas of peculiar kinematics in the collection of spectra 
taken with the smaller SB2 bundle that provides higher spectral 
resolution. We find many spectra in which the ionized lines split 
into two or more components with separations, in some cases, of
 more than 200 km/s. This line splitting was first reported by 
\cite{rubingraham90} and an analysis was afterwards performed by 
\cite{ceciletal92}, who detected a braided jet structure. 

\begin{figure}
   \centering
   \includegraphics[angle=0,width=8.5cm, clip=true]{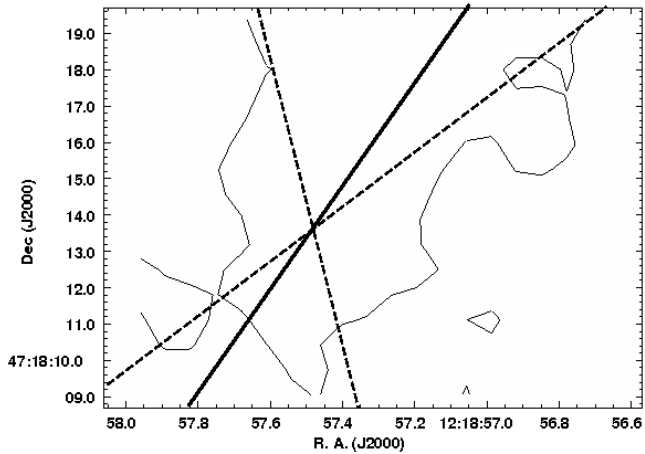}
   \caption{Contour of the regions where a second kinematic 
component can be found in the ionized gas. The dashed lines show a 
sketch of the tentative bipolar structure. The major axis is indicated.}\label{doscomps}
\end{figure}

In order to attempt the identification of different gaseous systems 
in the area covered by our observations, we have made a multi-component
 fit in the line of \cite{ceciletal92} based on the following assumptions: 
(i) there are two main kinematic systems of ionized gas, (ii) the 
broadening of the two components may be different, and (iii) the velocity of 
the two components is fairly different. Thus, we simultaneously fit the two 
components to the N[II], SII and H$\alpha$ emission lines. We accept the 
presence of two components when the goodness of the fit exceeds  the 
goodness of the fit to a single component by 30\%. Despite the intrinsic difficulties
 of this process and the peculiarities of this galaxy (e.g.\ in many cases 
the lines are not completely split, or there are obviously more than two 
components) which prevent the determination of reliable velocity maps for 
both kinematic components (as was achieved by \cite{Jimenezetal07} for a 
different object), it has been possible to identify consistently two main 
kinematic components at many locations throughout the FOV. The kinematics of the first one, 
which we identify with the gas in the galaxy disc,are not very
 different from that of the stars. The other component, shows  most peculiar,
 nearly counter-rotating kinematics. The locations of the spectra where this 
anomalous component is identified are represented in Figure \ref{doscomps}. We 
have chosen not to show the intensities, but only a sketch of the 
locations where two distinct kinematic components have been distinguished. 
Intensities and velocities in both components may be seriously affected by the difficulties mentioned above, but
locations where there is more than one component can be more reliably detected.
It must be pointed out that there still might be some locations where there are two
or more components which have not been detected by our procedure, particularly 
in the more extinguished region to the south. 
This figure shows that this perturbed ionized gas component has a bipolar 
morphology. The northern region is somewhat inclined westwards, roughly between 
the (nearly vertical) axis defined by the jet and the major axis, and the 
southern part, is roughly eastwards from the (N--S) jet axis.
The existence of a southern counterpart for the {\em northern Loop} as mentioned 
in Section \ref{morphokin} is also hinted by this figure.

\begin{figure}
   \centering
   \includegraphics[angle=0,width=8.5cm, clip=true]{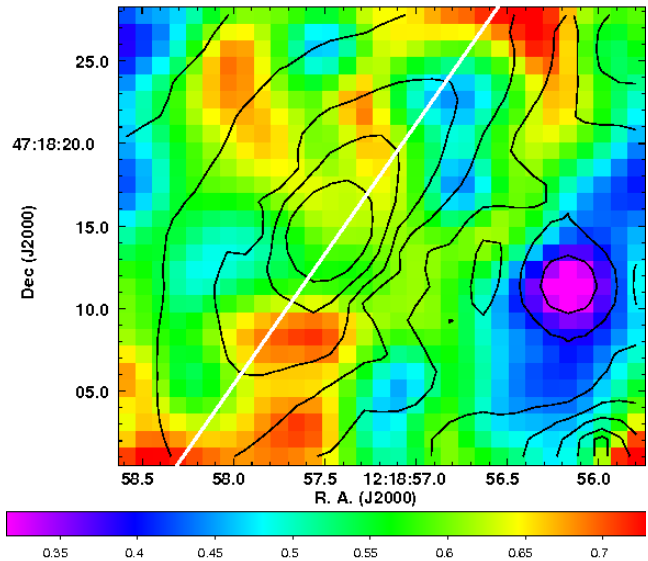}
   \caption{Map of the [NII]/H$\alpha$ ratio with the SB3 bundle. 
Contours show synthetic image with the F658N filter. The major axis is indicated by the white line.}\label{niioverha}
\end{figure}

\begin{figure}
   \centering
   \includegraphics[angle=0,width=8.5cm, clip=true]{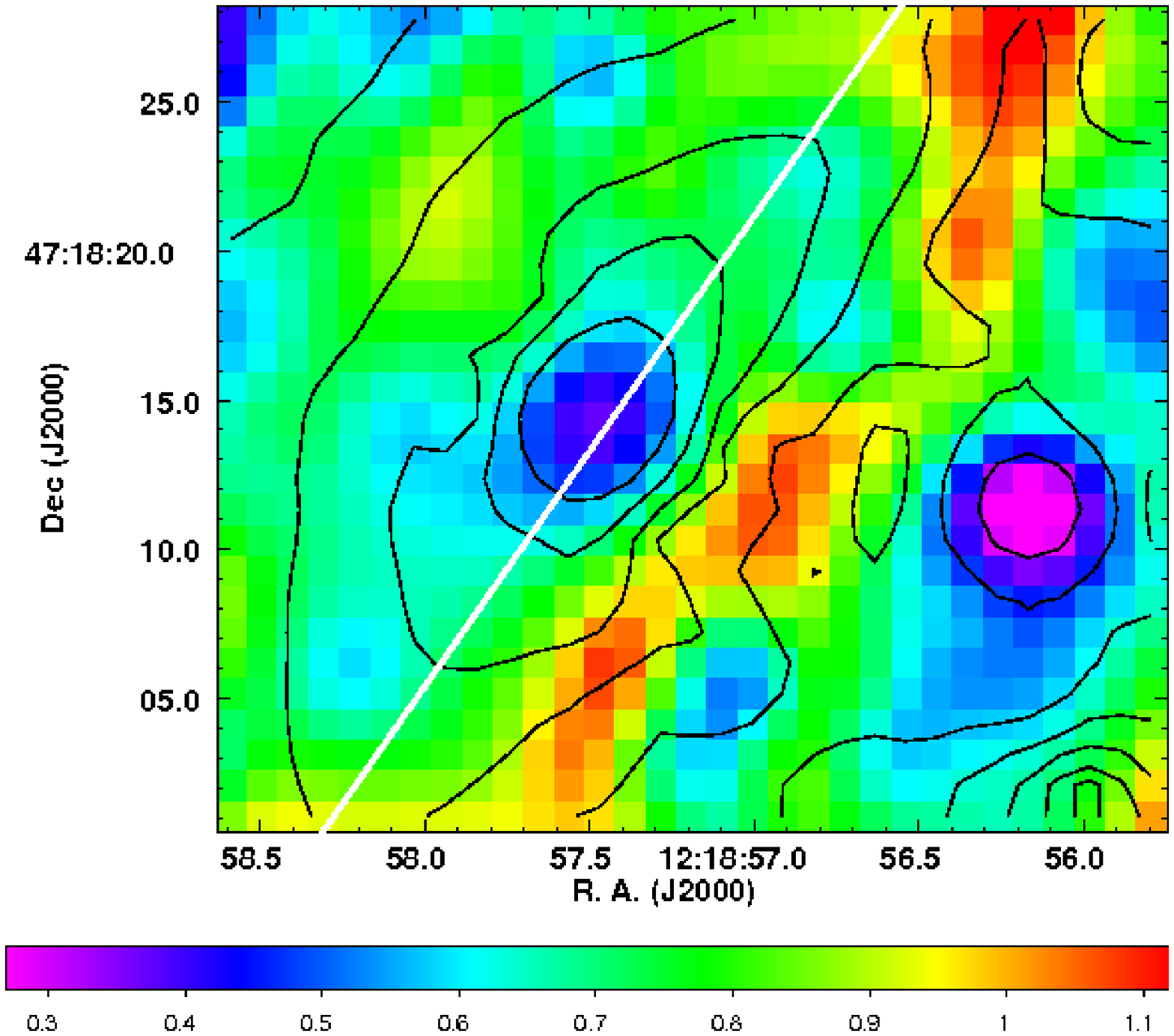}
   \caption{Map of the [SII]/H$\alpha$ ratio. Contours show synthetic 
image with the F658N filter. The major axis is indicated by the white line.}\label{siioverha}
\end{figure}

\begin{figure}
   \centering
   \includegraphics[angle=0,width=8.5cm, clip=true]{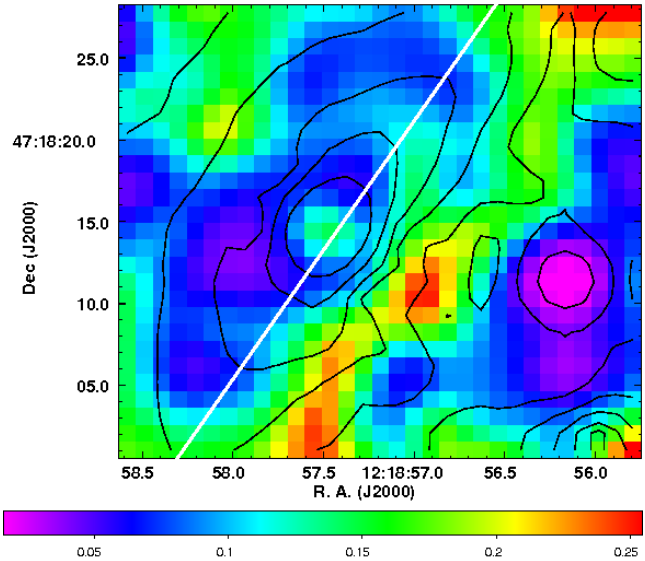}
   \caption{Map of the OI/H$\alpha$ ratio. The contours show the synthetic
 image with the F658N filter. The major axis is indicated by the white line.}\label{oioverha}
\end{figure}

In Figs. \ref{niioverha}, \ref{siioverha} and \ref{oioverha} 
the maps corresponding to the [NII]/H$\alpha$, [SII]/H$\alpha$
 and OI/H$\alpha$ emission line ratios are presented.  According
 to these maps the central region has relatively low [NII]/H$\alpha$, 
[SII]/H$\alpha$ and OI/H$\alpha$ emission line ratios. Taking
into account the high [OIII]/H$\beta$ measurements of \cite{ceciletal95}, 
our maps are consistent with photoionization as the main 
ionizing source for the central region, most probably induced by the 
Seyfert nucleus.  Surrounding this central region towards the west
 there are several regions of very high [SII]/H$\alpha$ and OI/H$\alpha$ 
emission line ratios forming an elongated strip almost parallel to the 
major geometric axis. There is also a region with high ratios on the
 other side of the nucleus, suggesting a ring-like structure for the 
shocked gas. Shocks are the main cause of ionization of the gas in this 
strip/ring (front of shock). The border of the ionized gas component with 
peculiar kinematics overlaps with this shock front in the NW.
Although shock-excited gas has been previously reported
 (cf.\ \cite{rubingraham90} or \cite{ceciletal95}) we have been able to map the
distribution of this shock ionized gas.

\section{Discussion}

Among the variety of models proposed to describe the peculiar features 
of NGC 4258 that we have mentioned in Section \ref{intro}, we  will 
discuss two: (a) the model described by \cite{wilsonetal01}, which 
explains the anomalous arms as dense gas in the disc shocked by gas driven 
onto the disc by the out-of-plane cocoon generated by the jets, and which 
seems favoured by the latest X-ray observations (cf.\ \cite{Sawadasatohetal07}); 
and (b) the bipolar hypershell model as described by \cite{SofueandVogler01}.
 Both models may be supported by the present observations.

\subsection{Jet-driven shocks in the galaxy disc}

In Fig. \ref{doscomps} we can see that the bipolar component of the ionized gas 
with peculiar kinematics lies roughly in the region between the N--S direction 
defined by the jet axis and the geometric major axis. This is the expected 
morphology in the scenario in which a cocoon of gas driven out of the plane by the 
jet is being blasted against the ISM disc (see \cite{wilsonetal01} and references
 therein). The bipolar component is mainly photoionized (most probably by the 
Seyfert nucleus) but its western border overlaps with the front shock detected 
in the [SII]/H$\alpha$ (see Fig. \ref{siioverha}). This supports the notion that photoionized 
gas driven by the jet component is shocking with the ISM disc forming a hot 
cocoon/interface.  According to this model, the front shock could trace the inner
 part of the anomalous arms.  To complete the picture, the observed 
kinematics of the bipolar component of ionized gas (towards the blue in the 
NW and towards the red in the SE) will fit in this scheme if the cocoon/interface 
is back-flowing.  Qualitatively, the presence of radial outflows in the neutral 
gas blasted by the cocoon would also support this model.

\subsection{The bipolar hypershell model}

\cite{Sofue80} and \cite{SofueandVogler01} interpreted the anomalous
 arms as the enhanced emission from the edges of a dumbbell-shaped shell 
that is symmetric with respect to the plane of NGC 4258. The presence of
 a  wind creating this structure is strongly supported by the detection
 of a very strong blueshift of the neutral gas emitters. 
This result is commonly accepted as solid evidence for a galactic wind, 
which is also supported by the existence of ionized gas systems of peculiar 
kinematics and the front shock. The main objection to Sofue's bipolar hypershell 
model (\cite{valbadavhulst82}) is the lack of symmetry of the  hypershells. 
Although in Sofue's model the hypershells are generated by a starburst, the 
emission line ratios indicate that in NGC 4258 the entraining gas could be 
photoionized by the AGN. According to \cite{veill05} when the source powering 
the galactic wind is an AGN, asymmetries in the expanding shell are expected. 
This is particularly true in the case of NGC 4258, where the out-of-plane jet
 would intersect the bipolar hypershell in the NS (projected) direction. The 
interaction between jet and hypershell would destroy that part of the hypershell 
bubbles (or would prevent it to form if the jet was in action when the shell was 
developing). Therefore, in this scenario, the anomalous arms would be the surviving 
parts of the hypershell.
Indeed, the morphology of the radio emission as shown in \cite{valbadavhulst82} 
and \cite{ceciletal00} strongly suggests this interpretation. The sharp outer
edges of the anomalous arms with the diffuse emission on the inside is
naturally explained in this scenario. The observational fact that the SE arm
is behind the galaxy disc while the NW arm is in front of it (cf.\ \cite{wilsonetal01}) 
also fits very naturally into this model. Finally, the nearly 
straight morphology of the radio emission in the innermost region shows that 
the jet is clearly dominant in this region, in contrast with the outermost
region, which is shell dominated.

\section{Conclusions}

We have performed two-dimensional spectroscopic observations in the central region
of NGC 4258 and have used these new data to test the available models for this
peculiar object. 
The main results derived from the analysis of these new 2D spectroscopy observations of NGC 4258 are:

\begin{enumerate}

\item The interstellar NaD absorption lines are strongly blueshifted.
 Comparison of the 2D velocity fields of the NaD with the stellar velocity 
field reveal that the blueshifted velocities are quite high, in the 
range $-350$ km/s to $-120$ km/s, and occur everywhere in the detected neutral gas 
on the near side of the galaxy disc (that is, the one that we can observe in absorption).

\item The velocity map of the ionized gas appears to be strongly distorted. 
There is a component of ionized gas of bipolar morphology with  peculiar, 
almost counter-rotating, kinematics.

\item The [SII]/H$\alpha$ and OI/H$\alpha$ emission line ratio maps show a 
thin ring-like region of high values indicative of shocks surrounding the 
galaxy nucleus. The border of the bipolar ionized gas component overlaps this 
front shock on the near side of the galaxy.

\end{enumerate}

The geometry of the bipolar ionized gas component and its partial overlapping 
with the front shock could fit into the model very well, in that hot gas driven
 by the jet blasts the ISM disc, giving rise to the anomalous arms. However, 
the radial motions of the ISM seem  hard to explain by this model alone. 
This outflow, could be naturally interpreted in terms of the presence of the 
bipolar hypershell model (\cite{Sofue80} and \cite{SofueandVogler01})  in that 
the anomalous arms would be the enhanced emission from the edges of the shells. 
The lack of symmetry of the bubbles (lack of presence of the anomalous arms in 
the NE and SW) would be explained by the disruption caused by the jet in the regions where it 
intersects the bubbles.

\section*{Acknowledgments}

We are grateful to the anonymous referee for his valuable comments which have
led to an improvement in this paper.
This paper has been supported by the ``Secretar\'{\i}a de
Estado de Pol\'{\i}tica Cient\'{\i}fica y Tecnol\'ogica'' 
(AYA2007-67625-C02-02) and by the ``Consejer\'{\i}a de Ciencia y T
ecnolog\'{\i}a de la Junta de Andaluc\'{\i}a'' (FQM-108, P05-FQM-792). A. 
Castillo-Morales acknowledges support from the Juan de la Cierva Programme financed 
by the Spanish MICINN and from the Spanish Programa Nacional de Astronom{\'i}a y 
Astrof{\'i}sica under grant AYA2006-02358.
This research has made use of NED
which
is operated by the JPL, Caltech, under contract with NASA. We have used observations
 made with the NASA/ESA HST, obtained from the data archive at the STScI.
STScI is operated by the Association of Universities for Research in
Astronomy, Inc. under NASA contract NAS 5-26555.

\end{document}